\documentclass[]{iopart}
\usepackage{graphicx,amssymb}
\usepackage[active]{srcltx}
\begin{document}
\title{Jammed disks in narrow channel: criticality and ordering tendencies}  

\author{Norman Gundlach$^{1}$, Michael Karbach$^{1,2}$, Dan Liu$^{2}$ and
  Gerhard M{\"{u}}ller$^{2}$} 

\address{$^{1}$ Bergische Universit{\"{a}}t Wuppertal,
  Fachbereich C, 42097 Wuppertal, Germany}

\address{$^{2}$ Department of Physics, University of Rhode Island, Kingston RI
  02881, USA}

%
\begin{abstract}
  A system of identical disks is confined to a narrow channel, closed off at
  one end by a stopper and at the other end by a piston.  All surfaces are
  hard and frictionless.  A uniform gravitational field is directed parallel
  to the plane of the disks and perpendicular to the axis of the channel.  We
  employ a method of configurational statistics that interprets jammed states
  as configurations of floating particles with structure.  The particles
  interlink according to set rules.  The two jammed microstates with smallest
  volume act as pseudo-vacuum.  The placement of particles is subject to a
  generalized Pauli principle.  Jammed macrostates are generated by random
  agitations and specified by two control variables.  They are inferred from
  measures for expansion work against the piston, gravitational potential
  energy, and intensity of random agitations.  In this two-dimensional space
  of variables there exists a critical point.  The jammed macrostate realized
  at the critical point depends on the path of approach.  We describe all
  jammed macrostates by volume and entropy.  Both are functions of the average
  population densities of particles.  Approaching the critical point in an
  extended space of control variables generates two types of jammed
  macrostates: states with random heterogeneities in mass density and states
  with domains of uniform mass density.  Criticality is shown to be robust
  against some effects of friction.
\end{abstract}


%
\section{Introduction}\label{sec:intro}
%
Granular matter consists of particles with sizes too big for thermal
fluctuations to have a significant impact.  The spatial configurations of
grains depend on many factors including their sizes and shapes, interactions
between them, interactions with external fields, and a protocol for producing
macrostates in a controllable way \cite{JNB96, dG99, NKB+98,Xu11}.  Models of
granular matter tend to omit non-essential attributes of grains in efforts to
gain more profound knowledge of bulk behavior.

In this spirit, jammed states of rigid objects with no interactions other than
hard-core repulsion have become the focus of numerous studies.  It has become
common to describe jammed states of granular matter in the language of
equilibrium statistical mechanics albeit with provisos.  We shall use the term
\emph{configurational statistics} for this framework of analysis \cite{EO89,
  ME89, EM94}.  Jammed states are frozen, hence time averages useless.
Nevertheless, jammed macrostates are postulated to exist, to be systematically
reproducible by means of some protocol of random agitations, and to be
describable by entropies and averaged mass densities in spatial regions of
macroscopic size.

The concepts of criticality, ordering, and transitions as used in studies of
granular matter have a dynamic context for the most part such as in
self-organized criticality \cite{BTW88} and jamming transitions
\cite{OLL+02,MSLB07,Head09,SNRS07}.  In this work we use these concepts in a
more restricted sense, where a critical singularity and ordering tendencies
are evident in the configurational entropy of jammed macrostates.  Our study
is focused on a simple scenario that enables us to investigate these phenomena
analytically.

\begin{figure}[!hbt]
  \begin{center}
 \includegraphics[width=65mm]{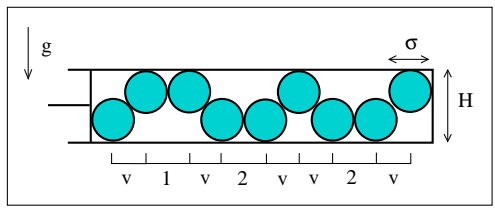}
\end{center}
\caption{Jammed microstate of disks of mass $m$ and diameter $\sigma$ in a
  channel of width $H$.  For $1<H/\sigma<1+\sqrt{3/4}$ all configurations are
  sequences of four distinct interlinking two-disk tiles.  In our methodology,
  two tiles (marked \textsf{v}) are elements of pseudo-vacuum and the other
  two (marked \textsf{1},\textsf{2}) are particles.}
  \label{fig:jamconf}
\end{figure}

Inspired by prior work \cite{BS06,AB09,BA11}, we consider a long, narrow
channel of width $H$ with the axis in a horizontal direction, a stopper at one
end, and a piston at the other end as shown in Fig.~\ref{fig:jamconf}.  The
channel contains $N$ disks of mass $m$ and diameter $\sigma$.  All surfaces
are rigid and frictionless.  Jamming requires three points of contact on each
disk that are not on the same semicircle.  A uniform gravitational field $g$
in the direction shown is present.

In channels with $1<H/\sigma<1+\sqrt{3/4}$, every jammed disk (except the two
outermost) touches one wall and two adjacent disks.  Every possible jammed
microstate is a sequence of interlinking two-disk tiles.  The four distinct
tiles are identified in Fig.~\ref{fig:jamconf}.  The two most compact states
consist of tiles \textsf{v} in alternating sequence. There is no need for
using two \textsf{v}-tile symbols.  Both tiles have the same volume and
energy.  One cannot be exchanged for the other in any given location.  The
least compact states are sequences $[\mathsf{v1v2v1v2}\cdots\mathsf{v}]$ and
$[\mathsf{v2v1v2v1}\cdots\mathsf{v}]$.  Only tiles \textsf{v} can follow each
other directly.  Any tile \textsf{1} (\textsf{2}) is separated from the next
tile \textsf{1} \textsf({2}) by at least two tiles \textsf{v}.  A tile
\textsf{1} is separated from the nearest tile \textsf{2} by at least one tile
\textsf{v}.

In channels with $1+\sqrt{3/4}<H/\sigma<2$, there exist 32 tiles interlinking
according to more complicated rules \cite{AB09}.  If $H/\sigma>2$, disks can
pass one another with the consequence that the jamming condition becomes
nonlocal.  Here we assume that $H/\sigma<1+\sqrt{3/4}$.

Jammed microstates in this system are countable.  Each has a well defined
volume.  Many have the same volume.  They represent one jammed macrostate in
the sense of an ensemble average.  The multiplicity of microstates with given
volume determines the configurational entropy of that macrostate.  In the
limit $N\to\infty$, the entropy per disk, $\bar{S}\doteq S/N$, becomes a
smooth function of reduced excess volume, $\bar{V}\doteq (V-V_0)/N$, where
$V_0$ is the volume of the most compact state.

The inverse slope of the configurational entropy curve is an intensive
variable named \emph{compactivity} \cite{EO89}:
\begin{equation}\label{eq:1}
X\doteq\left(\frac{d\bar{S}}{d\bar{V}}\right)^{-1}.
\end{equation}
This definition works well when a jammed macrostate is specified by a single
intensive variable.  Typical for such cases is that $\bar{S}$ increases
monotonically between $\bar{V}=0$ and some value $\bar{V}=\bar{V}_\infty$.
The function $\bar{S}(\bar{V})$ is concave along that stretch with infinite
initial slope and zero final slope.  The compactivity thus varies between
$X=0$ at $\bar{V}=0$ and $X=\infty$ at $\bar{V}=\bar{V}_\infty$.  It is a
measure for the intensity by which the system is randomly agitated to produce
a specific jammed macrostate.  A dosage of intense random agitations produces
a highly compactifiable state with much excess volume.  Jolting the system
with less intensity produces a more compact and less compactifiable jammed
macrostate.

Between the most disordered jammed macrostate at $\bar{V}=\bar{V}_\infty$ and
the least compact jammed macrostate at $\bar{V}=\bar{V}_{max}$ the function
$\bar{S}(\bar{V})$ stays concave and thus decreases.  These jammed macrostates
have negative compactivity.  They do exist but cannot, in general, be
generated by random agitations.

This scenario is borne out if we set $g=0$ in our system or tilt the channel
into a horizontal plane.  The configurational entropy of this case was
determined in \cite{BS06}.  The gravitational field $g$ complicates matters
significantly.  Jammed macrostates are now specified by two independent
intensive variables.  Microstates with equal volume have different statistical
weight on account of their gravitational potential energy.

We introduce a methodology that can cope with this situation
(Sec.~\ref{sec:meth}).  $\bar{S}$ turns out to be a multiple-valued function
of $\bar{V}$ (Sec.~\ref{sec:conf-entr}).  There exists a critical point
(Sec.~\ref{sec:crit}).  Ordering tendencies at criticality are explored
(Sec.~\ref{sec:domains}).  The analysis is extended to include some effects of
friction (Sec.~\ref{sec:fric}).  More complex scenarios now appear within
analytic reach (Sec.~\ref{sec:con}).

%
\section{Methodology}\label{sec:meth}
%
Here we adapt a method of statistical mechanical analysis based on fractional
exclusion statistics to jammed states of granular matter.  This approach was
originally developed for quantum many-body systems
\cite{Hald91a,Wu94,Isak94,Anghel} and was recently extended to classical
particles with shapes \cite{LVP+08,copic,picnnn,pichs}.  Whereas microscopic
degrees of freedom are subject to thermal fluctuations, the disks in the
narrow channel need to be subjected to random agitations to produce a
comparable effect.  In the context of this study the disks are not the
particles themselves.  The particles (or quasiparticles) are the tiles
\textsf{1} and \textsf{2}.  The method to be described has three parts.

\subsection{Energetics}\label{sec:ener}
Configurational statistics postulates that a protocol of random agitations
exists such that jammed macrostates are reproducibly generated.  Random
agitations of controlled intensity are performed while the piston exerts a
controlled force on the disks.  The disks are not jammed at this stage.  Their
dynamic state is influenced by competing agents: (i) the intensity of random
agitations, (ii) the force of the piston, and (iii) the weight of the disks.

A jammed microstate is frozen out when, abruptly and simultaneously or in
short order, the random agitations are stopped and the force of the piston is
increased to a value much larger than the weight of all disks combined. This
protocol is designed to keep the stable jammed microstates the same with or
without gravity.  Gravity merely affects their statistical weights.  The
associated jammed macrostate is then characterized by two ratios of three
intensive variables that reflect the three agents.  Each variable is a form of
energy competing for influence on the dynamic state and leaving its mark on
the jammed macrostate.

The volume of each element of pseudo-vacuum as indicated by a bracket in
Fig.~\ref{fig:jamconf} is $\sigma\sqrt{1-(H/\sigma-1)^2}$ and the volume of a
particle from either species is $\sigma$. Hence the activation of each
particle increases the volume by a fraction of the disk diameter,
\begin{equation}\label{eq:5} 
v_1=v_2=q\sigma,\quad q\doteq 1-\sqrt{1-(H/\sigma-1)^2}.
\end{equation}
To reduce clutter in the notation we use a channel of unit cross sectional
area and identify the force of the piston during random agitations with the
ambient pressure $p$. The activation of a particle \textsf{1} or \textsf{2}
adds energy.  One disk moves up or down in the gravitational field $g$ and the
piston moves out against pressure $p$:
\begin{equation}\label{eq:6} 
  \epsilon_{1}=pq\sigma+\gamma,\quad \epsilon_{2}=pq\sigma-\gamma,\quad
  \gamma=mg(H-\sigma).
\end{equation}
The third unit of energy that competes with $pq\sigma$ and $\gamma$ is some
measure $T_k$ for the intensity of random agitations.

For $g=0$ the jammed macrostate resulting from the random agitations at given
pressure only depends on the ratio
\begin{equation}\label{eq:3a} 
K_p\doteq\frac{pq\sigma}{T_k},
\end{equation}
which is a dimensionless version of $1/X$. In a thermal context we have
$(\partial\bar{S}/\partial\bar{V})_{\bar{U}}=p/T$. For $g\neq0$ the jammed
macrostate also depends on a second dimensionless ratio,
\begin{equation}\label{eq:3b} 
K_g\doteq\frac{\gamma}{T_k}.
\end{equation}
Configurational statistics does not specify how the intensity of random
fluctuations is quantitatively reflected in $T_k$.  This poses no problem.
Whereas the entropy $\bar{S}$ and the excess volume $\bar{V}$ are functions of
$K_p$ and $K_g$, the shape of the function $\bar{S}(\bar{V})$ inferred from
them only depends on
\begin{equation}\label{eq:4} 
\Gamma\doteq\frac{K_g}{K_p}=\frac{\gamma}{pq\sigma}.
\end{equation}

\subsection{Combinatorics}\label{sec:comb}
For the combinatorial analysis we use the template designed for statistically
interacting particles familiar from previous work (adapted to open boundary
conditions) \cite{Hald91a, copic}:
\begin{eqnarray}
  \label{eq:7a} 
  W(\{N_m\}) =n_{pv}\prod_{m=1}^M
  \left(\begin{array}{c}d_m+N_m-1 \\ N_m\end{array}\right), 
  \\ \label{eq:7b} 
  d_m =A_m-\sum_{m'=1} ^Mg_{mm'}(N_{m'}-\delta_{mm'}).
\end{eqnarray}
This multiplicity expression yields the number of microstates with given
particle content.  The degeneracy of the pseudo-vacuum is encoded in $n_{pv}$.
The $A_m$ are capacity constants and the $g_{mm'}$ are statistical interaction
coefficients.

All particles of a given species will have the same volume and energy no
matter where they are placed.  Therefore, all microstates with given particle
content constitute a macrostate in the sense discussed earlier.  The entropy
of a macrostate as derived from the multiplicity expression (\ref{eq:7a}) for
$N_m\gg1$ via $S=k_B\ln W$ reads \cite{Isak94, picnnn}:
\begin{eqnarray}\label{eq:8a}\hspace*{-15mm}
S(\{N_m\}) = k_B\sum_{m=1}^M
\Big[\big(N_{m}+Y_m\big)\ln\big(N_m+Y_m\big) 
-N_m\ln N_m -Y_m\ln Y_m\Big], 
\\ \label{eq:8b}
 Y_m \doteq A_m-\sum_{m'=1}^Mg_{mm'} N_{m'}.
\end{eqnarray}

\subsection{Statistical mechanics}\label{sec:stat}
The statistical mechanical analysis as transcribed from a thermal system
\cite{Wu94, Isak94} produces, for the grand partition function, the result
\cite{copic}
\begin{equation}\label{eq:9} 
Z=\prod_{m=1}^M\big(1+w_m^{-1}\big)^{A_m},
\end{equation}
where the (real, positive) $w_m$ are the solutions of the coupled nonlinear
algebraic equations,
\begin{equation}\label{eq:10} 
e^{K_m}=(1+w_m)\prod_{m'=1}^M \big(1+w_{m'}^{-1}\big)^{-g_{m'm}},
\end{equation}
and where $K_m\doteq\epsilon_m/T_k$.  The average population densities,
$\bar{N}_m\doteq \langle N_m\rangle/N$, of particles from each species are
derived from the coupled linear equations,
\begin{equation}\label{eq:11} 
w_m\bar{N}_m+\sum_{m'=1}^Mg_{mm'}\bar{N}_{m'} =\bar{A}_m,
\end{equation}
where $\bar{A}_m\doteq A_m/N$.  The configurational entropy,
$\bar{S}(\bar{V})$, follows from a scaled version of the entropy expression
(\ref{eq:8a}) in combination with the volume expression,
\begin{equation}\label{eq:12} 
\bar{V}(\{\bar{N}_m\})=\sum_{m=1}^M\bar{N}_mv_m.
\end{equation}
The function $\bar{S}(\bar{V})$ can also be derived from (\ref{eq:9}) via
$\bar{S}=-(\partial\bar{G}/\partial T)_p$ and
$\bar{V}=(\partial\bar{G}/\partial p)_T$, where
$\bar{G}=-\lim_{N\to\infty}N^{-1}k_BT\ln Z$ if we set $T_k=k_BT$ as in a
thermal context.

%
\section{Configurational entropy}\label{sec:conf-entr}
%
All jammed microstates such as the one shown in Fig.~\ref{fig:jamconf} contain
particles from $M=2$ species.  The specifications needed for the statistical
mechanical analysis are listed in Table~\ref{tab:1}.  In the taxonomy of
\cite{copic} both species are compacts.

\begin{table}
  \caption{Specifications of the particles represented by tiles \textsf{1} and \textsf{2}:
    species, volume and energy (relative to vacuum), capacity constant (left), and 
    interaction coefficients (right). The pseudo-vacuum has degeneracy $n_{pv}=2$.
    Its elements have (absolute) volume
    ${(1-q)\sigma}$ and energy $p(1-q)\sigma$.}\label{tab:1} 
  \begin{indented}
    \item[]{}
    \begin{tabular}{cccc}\hline\hline \rule[-2mm]{0mm}{6mm}
      $m$ & $v_m$ & $\epsilon_m$ & $A_{m}$ 
      \\ \hline \rule[-2mm]{0mm}{7mm}
      $1$ & ~~$q\sigma$ & ~~$pq\sigma+\gamma$ & ~~$\frac{1}{2}(N-3)$
      \\ \rule[-2mm]{0mm}{6mm}
      $2$ & ~~$q\sigma$ & ~~$pq\sigma-\gamma$ &
      ~~$\frac{1}{2}(N-3)$ \\ \hline\hline
    \end{tabular}\hspace{7mm}
    \begin{tabular}{c|rr} \hline\hline  \rule[-2mm]{0mm}{6mm}
      $g_{mm'}$ & ~~$1$ & ~~~$2$  \\ \hline \rule[-2mm]{0mm}{6mm}
      $1$ & $\frac{3}{2}$ & $\frac{1}{2}$ \\ \rule[-2mm]{0mm}{6mm}
      $2$ & $\frac{1}{2}$ & $\frac{3}{2}$ \\ \hline\hline 
    \end{tabular}
  \end{indented}
\end{table} 

\subsection{Zero gravity}\label{sec:zero}
In the special case $g=0$, particles \textsf{1} and \textsf{2} have equal
energy.  No distinction is necessary.  They can be merged into a single
species $\bar{1}$ with specifications $A_{\bar{1}}=N-3$ and
$g_{\bar{1}\bar{1}}=2$.  It is a type-1 merger in the classification of
\cite{pichs}.  The configurational entropy follows directly from (\ref{eq:8a})
with these specifications:
\begin{eqnarray}\label{eq:13}\hspace*{-15mm}
\bar{S}=k_B\big[(1-\bar{N}_{\bar{1}})\ln(1-\bar{N}_{\bar{1}})
-\bar{N}_{\bar{1}}\ln\bar{N}_{\bar{1}} 
-(1-2\bar{N}_{\bar{1}})\ln(1-2\bar{N}_{\bar{1}})\big],
\end{eqnarray}
where $\bar{V}=q\sigma\bar{N}_{\bar{1}}$.  This result was first derived in
\cite{BS06} by a different method.  In this case we do not have to solve
(\ref{eq:10}) and (\ref{eq:11}) to get to $\bar{S}(\bar{V})$.  Those solutions
are
\begin{equation}\label{eq:14} 
w_{\bar{1}}=\frac{1}{2}e^{K_{\bar{1}}}\left[1+\sqrt{1+4e^{-K_{\bar{1}}}}\right],
\quad \bar{N}_{\bar{1}}=\frac{1}{w_{\bar{1}}+2},
\end{equation}
where $K_{\bar{1}}=K_p$ from (\ref{eq:3a}).  The entropy inferred from
(\ref{eq:9}) reads
\begin{equation}\label{eq:15}
\bar{S}=k_B\left[\ln\big(1+w_{\bar{1}}^{-1}\big)+\frac{K_{\bar{1}}}{2+w_{\bar{1}}}\right].
\end{equation}
Expressions (\ref{eq:14}) and (\ref{eq:15}) are consistent with (\ref{eq:13}).
All attributes of $\bar{S}(\bar{V})$ described in Sec.~\ref{sec:intro} are
realized. Contact with quantities in Table I of \cite{EO89} are readily
established.  If we set $\tilde{X}\doteq K_{\bar{1}}^{-1}$,
$\tilde{V}\doteq\bar{V}/q\sigma$, $\tilde{S}\doteq\bar{S}/k_B$, and $w\doteq
w_{\bar{1}}$, implying $\tilde{V}=(2+w)^{-1}$, $e^{1/\tilde{X}}=w^2/(1+w)$,
$\tilde{S}=\ln(1+w^{-1})+\tilde{V}/\tilde{X}$, and $\tilde{Z}=(1+w^{-1})$, we
obtain $\tilde{Y}\doteq\tilde{V}-\tilde{X}\tilde{S}=-\tilde{X}\ln(1+w^{-1})$
and, hence, $\tilde{Z}=e^{-\tilde{Y}/\tilde{X}}$.

\subsection{Nonzero gravity}\label{sec:nonzero}
\begin{figure}[!hbt]
  \begin{center}
 \includegraphics[width=70mm]{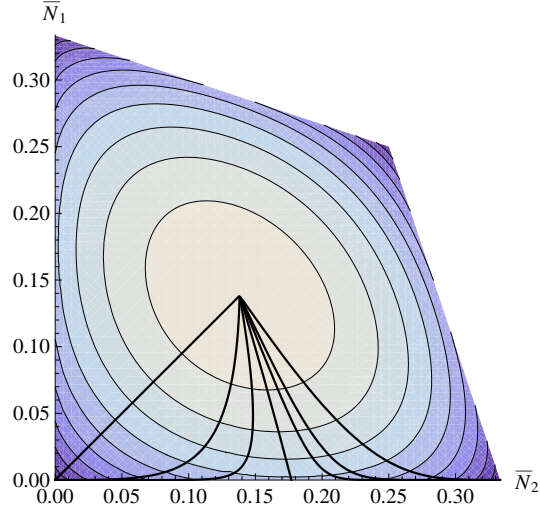}
\end{center}
\caption{Entropy per disk, $\bar{S}/k_B$, versus population densities
  $\bar{N}_1,\bar{N}_2$.  The contours are at $0.049\ell, \ell=1,\ldots,9$.
  The seven curves represent particle population densities of macrostates with
  ${\Gamma=0, 0.5, 0.75, 1, 1.25, 1.5, 3}$ (left to right).}
  \label{fig:cwlj2-1-f4}
\end{figure}

For $g\neq0$ the entropy depends on two independent variables.  The function
$\bar{S}(\bar{N}_1,\bar{N}_2)$ inferred from (\ref{eq:8a}) with the $A_m$ and
$g_{mm'}$ from Table~\ref{tab:1} produces an entropy landscape as shown in
Fig.~\ref{fig:cwlj2-1-f4}.  It is of quadrilateral shape with a smooth maximum
in the center and zeros at the corners.  All jammed states on a line
$\bar{N}_1+\bar{N}_2=\mathrm{const}$ have the same volume but different
gravitational potential energies.  Expression (\ref{eq:13}) is recovered by
$\bar{S}(\bar{N}_{\bar{1}}/2,\bar{N}_{\bar{1}}/2)$.  The system has the
highest capacity for particles if both species are present in equal numbers.

Equations (\ref{eq:10}) in simplified form read
\begin{equation}\label{eq:16} 
e^{2K_p}=\frac{w_1^2w_2^2}{(1+w_1)(1+w_2)},\quad e^{2K_g}=\frac{w_1}{w_2}.
\end{equation}
They reduce to a $4^{\mathrm{th}}$-order polynomial equation with one physically relevant,
real solution $w_m(K_p,K_g)$, $m=1,2$.
The particle population densities inferred from (\ref{eq:11}) are
\begin{eqnarray}\label{eq:17a} 
\bar{N}_1=\frac{w_2+1}{D},\quad  \bar{N}_2=\frac{w_1+1}{D}, 
\quad
D\doteq 2w_1w_2+3(w_1+w_2)+4.
\end{eqnarray}

Jammed macrostates generated by random agitations fall onto lines parametrized
by $\Gamma$.  Several such lines are shown in Fig.~\ref{fig:cwlj2-1-f4}.  All
lines begin at the jammed macrostate with maximum entropy and equal population
densities.  With $K_p$ increasing from zero the lines fan out.  On the line
$\Gamma=0$ the two population densities are being depleted at the same rate.
Higher values of $\Gamma$ have the effect that particles \textsf{1} are
depleted more rapidly.  The population of particles \textsf{2} may decrease or
increase as the curves show.

\begin{figure}[!hbt]
  \begin{center}
 \includegraphics[width=75mm]{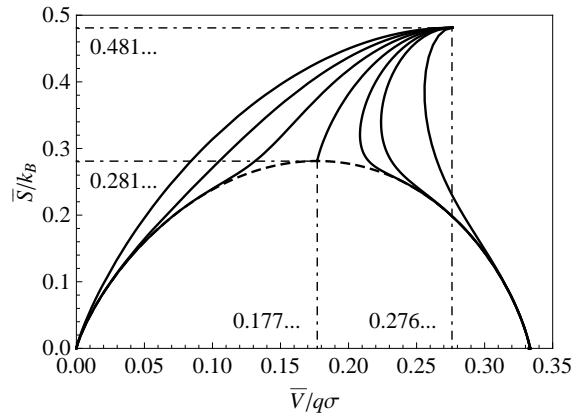}
\end{center}
\caption{Entropy per disk versus excess volume at $\Gamma=0, 0.5, 0.75,1,
  1.25, 1.5, 3$ (left to right).  The dashed arch delimits a region of
  critical macrostates.}
  \label{fig:cwlj2-1-f8}
\end{figure}

At $\Gamma=1$ particles 2 have zero activation energy.  Their density
increases moderately.  This is an entropic effect.  As the population of
particles 1 (with positive activation energy) decreases, the entropy is
maximized by an increasing population of particles 2.  The (nearly straight)
line pertaining to this case terminates at the point of maximum entropy for
states with $\bar{N}_1=0$.  The configurational entropy curves,
$\bar{S}(\bar{V})$, as inferred from (\ref{eq:16}) and (\ref{eq:17a}) in
combination with (\ref{eq:8a}) are shown in Fig.~\ref{fig:cwlj2-1-f8} for the
same values of $\Gamma$.

The concave curve on the upper left represents the zero-gravity case discussed
in Sec.~\ref{sec:zero} and previously solved in \cite{BS06}.  The inverse
slope of that curve is the compactivity (\ref{eq:1}), a quantity that varies
between zero in the most compact jammed macrostate and infinity in the
macrostate with the highest possible entropy.  All curves for nonzero gravity
start from that highest-entropy point and fan out in different directions.

At $\Gamma<1$ both species of particles have positive activation energies.
The physics does not change qualitatively from the zero-gravity case.  The
gravitational field suppresses the population density of particles \textsf{1}
and enhances that of particles \textsf{2}.  The net effect is a modest
reduction of entropy (of mixing) at given volume.  All curves end in the most
compact jammed state.

At $\Gamma>1$, the particles from species \textsf{2} have negative activation
energies.  The jammed state produced by random agitations in the low-intensity
limit is now qualitatively different.  It is a periodic array of close-packed
particles \textsf{2}.  It has volume $\bar{V}/q\sigma=\frac{1}{3}$, which is
larger than the ($\Gamma$-independent) jammed state produced in the
high-intensity limit of random agitations.

Increasing the intensity of random agitation always increases the entropy.
However, at $\Gamma>1$, it initially decreases the volume of the states thus
produced.  The decreasing $\bar{V}$ is associated with compression work used
for weight lifting.  The annihilation of particles \textsf{2} creates disorder
and thus increases $\bar{S}$.  At higher intensities the jammed states have
larger volume again.  Here particles \textsf{1}, which have high energy, are
created in significant numbers.

For the borderline case $\Gamma=1$ between the two regimes, the particles from
species \textsf{2} have zero activation energy.  Gravitational energy and
expansion energy are in balance.  Random agitations do not produce ordering in
the low-intensity limit.  The shapes of the curves in
Fig.~\ref{fig:cwlj2-1-f8} signal the presence of a critical singularity.

%
\section{Criticality}\label{sec:crit}
%
In the framework of configurational statistics it is legitimate to use the
term \emph{criticality} more loosely than is common in the theory of (thermal)
phase transitions.  In our system the critical singularity is associated with
the combined limit
\begin{equation}\label{eq:18} 
K_p\to\infty,\quad K_g\to\infty,\quad \Gamma\to1.
\end{equation}
The jammed macrostate realized at the critical singularity depends on how the
singularity is approached, which is not altogether unusual \cite{CCPZ12}.  
In Sec.~\ref{sec:conf-entr} we have considered one
particular approach, with $\Gamma=1$, implying that $w_1\leadsto\infty$ and
\begin{equation}\label{eq:31} 
w_2\leadsto\frac{(9-\sqrt{69})^{1/3}+(9+\sqrt{69})^{1/3}}{2^{1/3}3^{2/3}}=1.3247\ldots,
\end{equation}
the physically relevant solution of the cubic equation,
\begin{equation}\label{eq:19} 
w_2^3-w_2-1=0.
\end{equation}
The associated values for excess volume and entropy are%
\begin{eqnarray}\label{eq:32a} 
\bar{S}/k_B &\leadsto\ln w_2=0.28119\ldots,\\
\bar{V}/q\sigma &\leadsto(3+2w_2)^{-1}=0.17700\ldots.
\end{eqnarray}

Further critical macrostates are represented by points on the (dashed) arching
line in Fig.~\ref{fig:cwlj2-1-f8} and by points in the space underneath.  The
latter will be discussed in Sec.~\ref{sec:domains}.  The former are realized
in a family of pathways toward the critical singularity specified by a single
parameter,
\begin{equation}\label{eq:20} 
\Delta\doteq K_p-K_g.
\end{equation}
The critical macrostates are determined from
\begin{equation}\label{eq:21} 
w_1=\infty,\quad e^{2\Delta}=\frac{w_2^3}{1+w_2},\quad -\infty<\Delta<+\infty
\end{equation}
and the particle population densities (\ref{eq:11}) become
\begin{equation}\label{eq:22} 
\bar{N}_1=0,\quad \bar{N}_2=\frac{1}{3+2w_2}.
\end{equation}
The point $\Delta=0$ recovers the case $\Gamma=1$ discussed previously.  Only
particles \textsf{2} are present.  They are distributed randomly along the
channel.  The volume and the entropy depend on the population density
$\bar{N}_2$:
\begin{eqnarray}\label{eq:23a}\hspace{-15mm}
\bar{V}/q\sigma=\bar{N}_2, \\ \hspace{-15mm}
\bar{S}/k_B=\frac{1-\bar{N}_2}{2}\ln\left(\frac{1-\bar{N}_2}{2}\right)-\bar{N}_2\ln\bar{N}_2
-\frac{1-3\bar{N}_2}{2}\ln\left(\frac{1-3\bar{N}_2}{2}\right).
\end{eqnarray}
Complete ordering exists if there are no particles or if the particles are
close packed.  The function $\bar{S}(\bar{V})$ has zero slope at the point
(\ref{eq:32a}) and infinite slope at $\bar{S}=0$:
\begin{equation}\label{eq:33} 
\frac{d\bar{S}}{d\bar{V}}\leadsto \left\{
\begin{array}{ll} 
-\ln\bar{V}, & \bar{V}\to0 \\
\frac{3}{2}\ln\left(\frac{1}{3}-\bar{V}\right), & \bar{V}\to\frac{1}{3}
\end{array}\right..
\end{equation}

All critical macrostates are averages of the same subset of jammed
microstates, but weighted differently.  The macrostate at the top of the arch
has all microstates from that subset weighted equally.  The macrostates
elsewhere on the arch result from averages with one bias in the weighting.
The parameter $\Delta$ controls the population density $\bar{N}_2$ and thus
the volume.  All microstates with a given number of particles \textsf{2}
remain weighted equally.  The macrostates inside the arch result from averages
with two biases in the weighting.

%
\section{Ordering tendencies}\label{sec:domains}
%
We choose a second bias in the form of a weak interaction force that either
enhances or suppresses heterogeneities in mass density along the channel.  It
controls the degree of clustering of particles \textsf{2}.  New pathways to
the critical point generate macrostates with domains of uniform mass density.
No symmetry-breaking field needs to be introduced to make that happen.

We extend the set of particle species from two to three.  We split one species
into two: every compact \textsf{2} in the system becomes either a host
$\mathsf{2'}$ or a tag $\mathsf{2''}$ \cite{copic}.  The particle \textsf{2}
closest to the piston is a host $\mathsf{2'}$.  Any other particle \textsf{2}
is a tag $\mathsf{2''}$ if it follows the preceding particle \textsf{2} in one
of the two shortest tile sequences: \textsf{2vv2} or \textsf{2v1v2}.
Otherwise it is a host again.

In Fig.~\ref{fig:jamconf} we have one host $\mathsf{2}'$ followed by a tag
$\mathsf{2}''$ (left to right).  We do not split compacts \textsf{1} because
they become depleted as criticality is approached.  The specifications of the
three species, compact \textsf{1}, host $\mathsf{2}'$, and tag $\mathsf{2}''$,
are compiled in Table~\ref{tab:2}.  Anghel's rules \cite{Anghel,pichs} are
satisfied.

\begin{table}[htb]
  \caption{Specifications of three species of particles:
    species, volume and energy (relative to vacuum), capacity constant (left), and 
    statistical interaction coefficients (right).}\label{tab:2} 
  \begin{indented}
  \item[]
    \begin{tabular}{cccc}\hline\hline \rule[-2mm]{0mm}{6mm}
      $m$ & $v_m$ & $\epsilon_m$ & $A_{m}$ 
      \\ \hline \rule[-2mm]{0mm}{7mm}
      $1$ & $q\sigma$ & $pq\sigma +\gamma$ & $\frac{1}{2}(N-3)$
      \\ \rule[-2mm]{0mm}{6mm}
      $2'$ & $q\sigma$ & $pq\sigma-\gamma$ &
      $\frac{1}{2}(N-3)$ \\ \rule[-2mm]{0mm}{6mm}
      $2''$ & $q\sigma$ & $pq\sigma-\gamma-\phi$ &
      $0$ \\ \hline\hline
    \end{tabular}\hspace{5mm}
    \begin{tabular}{c|rrr} \hline\hline  \rule[-2mm]{0mm}{6mm}
      $g_{mm'}$ & ~$1$ & ~~$2'$ & ~~$2''$ \\ \hline \rule[-2mm]{0mm}{6mm}
      $1$ & $\frac{3}{2}$ & $\frac{1}{2}$ & $\frac{1}{2}$ \\ \rule[-2mm]{0mm}{6mm}
      $2'$ & $\frac{1}{2}$ & $\frac{5}{2}$ & $\frac{3}{2}$\\ \rule[-2mm]{0mm}{6mm}
      $2''$ & $0$ & $-1$ & $0$\\ \hline\hline 
    \end{tabular}
  \end{indented}
\end{table} 

Raising (lowering) the activation energy of the tag relative to that of the
host amounts to a repulsive (attractive) short-range force between particles
\textsf{2}.  Such a force suppresses (enhances) heterogeneities in mass
density.  It opens up new pathways to the critical singularity.  In the
critical macrostates thus generated the microstates with a given number of
particles \textsf{2} are no longer weighted equally.  The weighting now
discriminates between hosts and tags.

The evidence for the formation of domains presented here is indirect, as found
in the entropy. Spatial correlations will be the focus of a separate study.
Assembling the function $\bar{S}(\bar{V})$ in the extended parameter space
starts from the extended Eqs.~(\ref{eq:16}),
\begin{eqnarray}\label{eq:24} \fl
e^{2K_p}=\frac{w_1^2w_{2'}^3(1+w_{2''})}{(1+w_1)(1+w_{2'})^2w_{2''}},
\quad 
e^{2K_g}=\frac{w_1(1+w_{2'})w_{2''}}{w_{2'}^2(1+w_{2''})}, \quad e^{\Phi}=\frac{w_{2'}}{w_{2''}},
\end{eqnarray}
where $\Phi\doteq\phi/T_k$ is an amendment to (\ref{eq:3a}) and (\ref{eq:3b}).
The population densities of compacts \textsf{1}, hosts $\mathsf{2'}$, and tags
$\mathsf{2''}$ are
\begin{eqnarray}\label{eq:25} \hspace{-15mm}
\bar{N}_1=\frac{1+2w_{2''}+w_{2'}w_{2''}}{D^{(\phi)}}, \quad 
\bar{N}_{2'}=\frac{(1+w_1)w_{2''}}{D^{(\phi)}},\quad
\bar{N}_{2''}=\frac{1+w_1}{D^{(\phi)}}, 
\\ \hspace{-15mm}
D^{(\phi)}=4+3w_1+7w_{2''}+5w_1w_{2''} 
+3w_{2'}w_{2''}+2w_1w_{2'}w_{2''}. \nonumber
\end{eqnarray}

\begin{figure}[!hbt]
  \begin{center}
 \includegraphics[width=70mm]{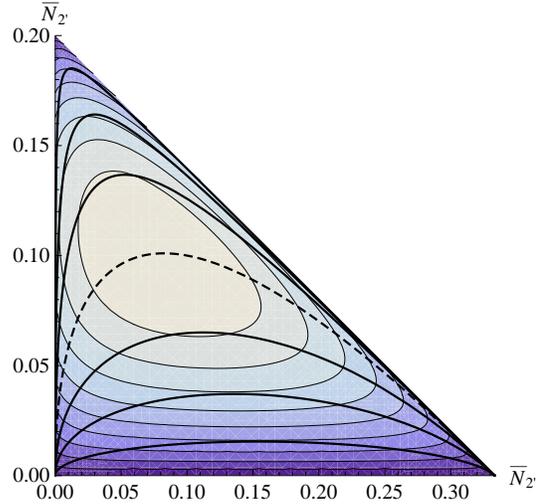}
\end{center}
\caption{Entropy per disk, $\bar{S}/k_B$, versus population densities
  $\bar{N}_{2''},\bar{N}_{2'}$ of particles from the two species $2''$ and
  $2'$. The contours are at $0.028\ell, \ell=1,\ldots,9$.  The solid and
  dashed curves represent (\ref{eq:29}) for $\Phi=0, \pm1.5, \pm3.0, \pm5.0$
  in ascending order from top to bottom.}
  \label{fig:janac1-4-f1}
\end{figure}

We now have a two-parameter family of pathways toward the critical singularity
with critical macrostates determined from
\begin{eqnarray}\label{eq:27} 
&w_1=0,\quad e^{2\Delta}=\frac{w_{2'}^5(1+w_{2''})^2}{(1+w_{2'})^3w_{2''}},
\quad e^{\Phi}=\frac{w_{2'}}{w_{2''}},  \\
&\hspace{5mm}-\infty<\Delta<+\infty,\quad -\infty<\Phi<+\infty, \nonumber
\end{eqnarray}
and with critical population densities,
\begin{equation}\label{eq:29} 
\bar{N}_1=0,\quad \bar{N}_{2'}=w_{2''}\bar{N}_{2''}=\frac{w_{2''}}{3+5w_{2''}+2w_{2'}w_{2''}}.
\end{equation}
The function $\bar{S}(\bar{N}_{2'},\bar{N}_{2''})$ inferred from (\ref{eq:8a})
is shown in Fig.~\ref{fig:janac1-4-f1}.  Every point in this entropy landscape
represents a macrostate at criticality.  All critical macrostates have
$pq\sigma=\gamma$ and $\phi=0$.

In macrostates pertaining to $\Phi=0$, particles \textsf{2} form random
clusters.  They are the states of highest entropy at given
$\bar{N}_{2'}+\bar{N}_{2''}$, represented by the dashed curve in
Fig.~\ref{fig:janac1-4-f1}.  This particular mix of hosts and tags,
\begin{equation}\label{eq:30} 
\bar{N}_{2'}=\frac{1}{4}\left[\sqrt{\bar{N}_{2''}(8+\bar{N}_{2''})}-5\bar{N}_{2''}\right],
\end{equation}
produces a characteristic texture of heterogeneity in mass density.  Critical
macrostates on either side of the dashed curve have mass heterogeneities with
textures that reflect a higher degree of ordering.

In macrostates generated on pathways with $\Phi>0$ tags are more abundant and
hosts less abundant than in a mixture of randomly placed particles \textsf{2}.
The bottom three solid curves are critical population densities (\ref{eq:29})
at fixed $\Phi>0$.  Clusters of hosts with tags are domains of uniform, low
mass density, phase separated from domains of uniform, high mass density.  The
latter are clusters of tiles \textsf{v}.

\begin{figure}[t]
  \begin{center}
    \includegraphics[width=63mm]{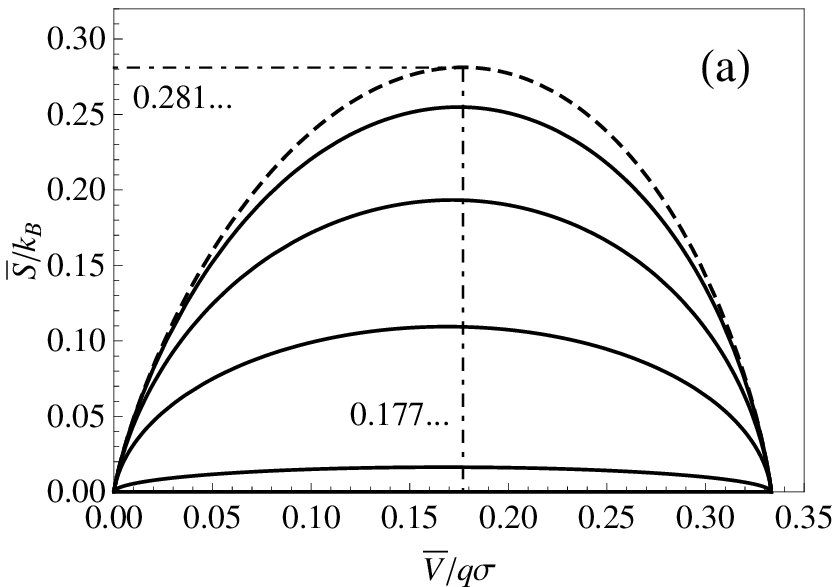}
    \includegraphics[width=63mm]{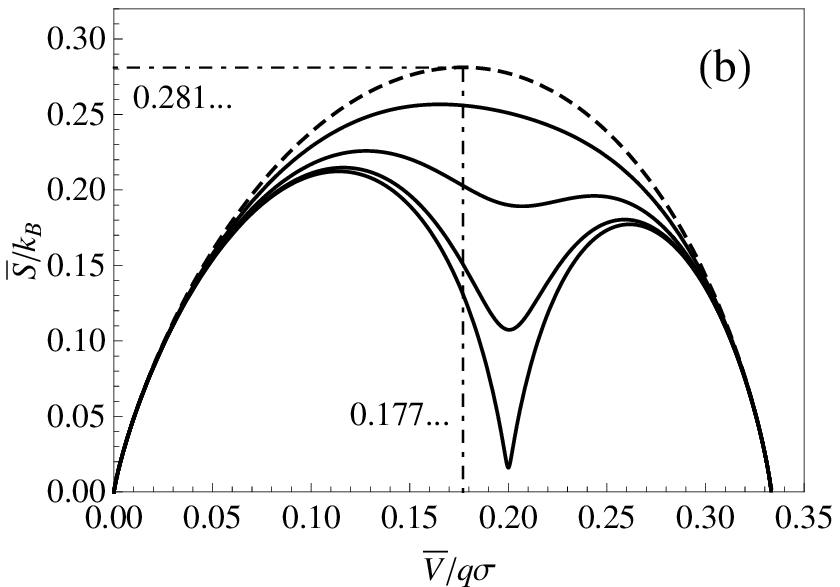}
\end{center}
\caption{Entropy per disk versus excess volume at criticality.  The five
  curves, inferred from (\ref{eq:27}), represent, from top to bottom, critical
  macrostates characterized (a) by $\Phi=0, 1.5, 3, 5, 10$. and (b) by
  $\Phi=0, -1.5, -3, -5, -10$}
  \label{fig:janac1-4-f3}
\end{figure}

These domains suppress the entropy significantly relative to the entropy of
random clusters of particles \textsf{2}.  The entropy reduction is evident in
the contours of Fig~\ref{fig:janac1-4-f1} and, more compellingly, in the
configurational entropy curves, $\bar{S}(\bar{V})$, shown in
Fig.~\ref{fig:janac1-4-f3}(a).  The effect is most significant at intermediate
volume where large clusters of tiles \textsf{2} or tiles \textsf{v} are least
likely to be realized by chance.

A different ordering tendency is manifest in critical macrostates generated
along pathways with $\Phi<0$.  More hosts and fewer tags are present than in a
randomly placed mixture.  Critical population densities (\ref{eq:29}) at fixed
$\Phi<0$ are represented by the top three solid curves in
Fig.~\ref{fig:janac1-4-f1}.  These macrostates exhibit various degrees of
dispersal of particles \textsf{2}.  The dispersal flattens out heterogeneities
in mass density and thus reduces the entropy.

This ordering tendency manifests itself differently in the configurational
entropy as shown in Fig.~\ref{fig:janac1-4-f3}(b).  Dispersal of particles
\textsf{2} only produces ordering tendencies in conjunction with spatial
constraints.  The entropy reduction remains insignificant until crowding of
particles \textsf{2} becomes an issue.  The capacity of the channel for hosts
$\mathsf{2'}$ is only 60\% of its capacity for particles \textsf{2}.  Hence,
hosts alone reach saturation at $\bar{V}/q\sigma=\frac{1}{5}$ in an ordered
state.  At larger $\bar{V}$, the entropy rises again on account of tags
attached to some hosts.


%
\section{Friction}\label{sec:fric}
%
The assumption of frictionless surfaces puts some distance between physical
reality and the results presented thus far.  Here we narrow that gap somewhat.
Friction will stabilize additional configurations of jammed disks.  However, a
rigorous analysis in the framework of configurational statistics now appears
remote.  The jammed microstates are no longer countable.  They include
configurations with disks that do not touch any wall.  Friction allows a disk
to be wedged in between two disks that touch the same wall.  The range of
distance between the middle disk and the wall depends on the coefficient of
friction.

It appears reasonable to argue that the dominant effect of friction is well
represented if we allow the controlled presence of adjacent disks that touch
the same wall.  Our method is readily adaptable to accommodate this scenario.
We extend the set of statistically interacting particles from two species
(Table~\ref{tab:1}) to four species (Table~\ref{tab:3}).  Any tile \textsf{1}
that is stable in the absence of friction remains a particle of species
\textsf{1}.  Any tile \textsf{1} that directly follows another tile \textsf{1}
can only be stabilized by friction.  We name it a particle \textsf{4}.
Likewise, in any string of adjacent tiles \textsf{2}, the first is a particle
\textsf{2} and the others are particles \textsf{3}.

\begin{table}[htb]
  \caption{Specifications of four species of particles:
  species, volume and energy (relative to vacuum), capacity constant (left), and 
  statistical interaction coefficients (right).}\label{tab:3} 
\begin{indented}
  \item[]
    \begin{tabular}{cccc}\hline\hline \rule[-2mm]{0mm}{6mm}
      $m$ & $v_m$ & $\epsilon_m$ & $A_{m}$ 
      \\ \hline \rule[-2mm]{0mm}{7mm}
      $1$ & $q\sigma$ & $pq\sigma +\gamma$ & $\frac{1}{2}(N-3)$
      \\ \rule[-2mm]{0mm}{6mm}
      $4$ & $q\sigma$ & $pq\sigma +\gamma+\psi$ & $0$
      \\ \rule[-2mm]{0mm}{6mm}
      $2$ & $q\sigma$ & $pq\sigma-\gamma$ &
      $\frac{1}{2}(N-3)$ \\ \rule[-2mm]{0mm}{6mm}
      $3$ & $q\sigma$ & $pq\sigma-\gamma+\psi$ &
      $0$ \\ \hline\hline
    \end{tabular}\hspace{5mm}
    \begin{tabular}{c|rrrrr} \hline\hline  \rule[-2mm]{0mm}{6mm}
      $g_{mm'}$ & ~$1$ & ~~$4$ & ~~$2$ & ~~$3$ \\ \hline \rule[-2mm]{0mm}{6mm}
      $1$ & $\frac{3}{2}$ & $\frac{1}{2}$ & $\frac{1}{2}$  & $\frac{1}{2}$\\ \rule[-2mm]{0mm}{6mm}
      $4$ & $-1$ & $0$ & $0$  & $0$ \\ \rule[-2mm]{0mm}{6mm}
      $2$ & $\frac{1}{2}$ & $\frac{1}{2}$ & $\frac{3}{2}$ & $\frac{1}{2}$\\ \rule[-2mm]{0mm}{6mm}
      $3$ & $0$ & $0$ & $-1$ & $0$\\ \hline\hline 
    \end{tabular}
  \end{indented}
\end{table} 

All four species of particles extend the volume by the same amount $q\sigma$.
Their expansion work is identical.  The gravitational potential energy is
raised (lowered) by $\gamma$ when a particle \textsf{1} or \textsf{4}
(\textsf{2} or \textsf{3}) is activated.  The population densities of
particles \textsf{4} and \textsf{3} relative to those of particles \textsf{1}
and \textsf{2}, respectively, are controllable by the parameter $\psi$ in a
role akin to a chemical potential. Unlike $pq\sigma$ in (\ref{eq:3a}) and
$\gamma$ in (\ref{eq:3b}), the parameter $\psi$ is not associated with any
physically relevant unit of energy. However, its value is determined, via
$R_f$, by an observable effect of friction. In the taxonomy of
Ref.~\cite{copic} particle \textsf{1} is host to tag particle \textsf{4} and
particle \textsf{2} host to tag \textsf{3}.

Equations (\ref{eq:10}) for the two pairs of hosts and tags become
\begin{eqnarray}\label{eq:2} 
e^{2K_p}=\frac{w_1^2w_2^2(1+w_3)(1+w_4)}{(1+w_1)(1+w_2)w_3w_4}, 
\nonumber \\
e^{2K_g}=\frac{w_1w_3(1+w_4)}{w_2(1+w_3)w_4}, \quad e^{\Psi}=\frac{w_4}{w_1}
=\frac{w_3}{w_2},
\end{eqnarray}
where $\Psi\doteq\psi/T_k$.
The particle population densities inferred from (\ref{eq:11}) are
\begin{eqnarray}\label{eq:26}\hspace{-10mm}
&\bar{N}_4 =\frac{\bar{N}_1}{w_4}=\frac{(1+w_2)w_3}{D^{(\psi)}},\quad 
\bar{N}_3 =\frac{\bar{N}_2}{w_3}=\frac{(1+w_1)w_4}{D^{(\psi)}}, \hspace{-10mm} \\
&D^{(\psi)} =(1+w_2)w_3+(1+w_1)w_4 +w_3w_4[2w_1w_2+3(w_1+w_2)+4]. \nonumber
\end{eqnarray}
In the limit $\psi\to\infty$ we have $w_3,w_4\to\infty$ and expressions
(\ref{eq:16}), (\ref{eq:17a}) for the frictionless case are recovered.

Within the limitations of our approach, the effects of friction are most
transparently accounted for if we impose the constraints
\begin{equation}\label{eq:28} 
\frac{\bar{N}_4}{\bar{N}_1}=\frac{\bar{N}_3}{\bar{N}_2}=R_f,
\end{equation}
because friction is the same near both walls. The parameter $R_f$ with range
$0\leq R_f\leq 1$ is a measure for the coefficient of friction. This simple
model to include friction can be modified in the light of evidence from
experiments or simulations.  We could prohibit, for example, the occurrence of
more than three tiles \textsf{1} or three tiles \textsf{2} adjacent to each
other. Within the limitations of this model we can state that $R_f$ is zero in
the absence of friction and that it increases monotonically with the
coefficient of static friction. The asymmetry in Eq.~(\ref{eq:28}) is due to
gravity, $\bar{N}_4<\bar{N}_3$ follows from $\bar{N}_1<\bar{N}_2$.

The four Eqs.~(\ref{eq:2}) are thus reduced to a pair,
\begin{equation}\label{eq:34} 
\frac{e^{2K_p}}{(1+R_f)^2}=\frac{w_1^2w_2^2}{(1+w_1)(1+w_2)},\quad e^{2K_g}=\frac{w_1}{w_2}.
\end{equation}
Expressions (\ref{eq:26}) for the particle population densities remain intact.

\begin{figure}[t]
  \begin{center}
 \includegraphics[width=75mm]{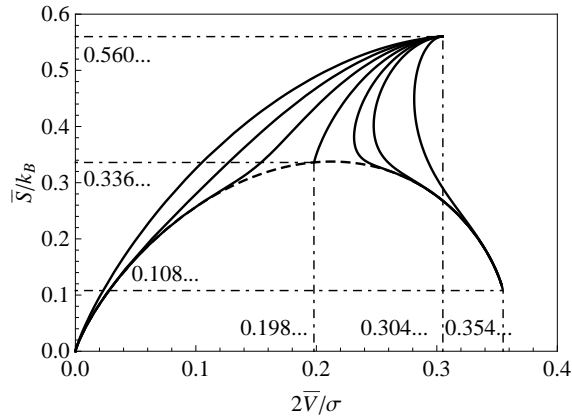}
\end{center}
\caption{Entropy per disk versus excess volume at $\Gamma=0, 0.5, 0.75,1,
  1.25, 1.5, 3$ (left to right) with effects of friction included $(R_f=0.1)$.
  The dashed arch delimits a region of critical macrostates.}
  \label{fig:six}
\end{figure}

How do the additional tiles stabilized by friction, i.e. the tag particles
\textsf{3} and \textsf{4}, affect the configurational entropy curves
$\bar{S}(\bar{V})$?  The answer is shown in Fig.~\ref{fig:six} for $R_f=0.1$.
In this case we have one tile \textsf{1} (\textsf{2}) stabilized by friction
for every ten tiles \textsf{1} (\textsf{2}) that are stable under pressure and
gravity alone.

All features except one are qualitatively the same as in
Fig.~\ref{fig:cwlj2-1-f8} for the frictionless case.  The curves fan out from
the most disordered state in a similar manner.  Criticality is robust.  The
coordinates of three landmarks (critical point, points of maximum and minimum
volume) have shifted somewhat.  The jammed macrostate with maximum volume now
has a nonzero entropy.  The source of this entropy are the extra tiles
stabilized by friction.  They are randomly distributed as tags \textsf{3} to
hosts \textsf{2}.  Tiles 1 are absent from this macrostate.

The critical states, located on the dashed curve in Fig.~\ref{fig:six}, are
again associated with the combined limit (\ref{eq:18}).  Equation
(\ref{eq:21}) for $w_2$ remains valid if we divide the left-hand side by
$(1+R_f)^2$.  The critical particle population densities in the presence of
friction are
\begin{equation}\label{eq:35} 
\bar{N}_1=\bar{N}_4=0,\quad \bar{N}_2=\frac{\bar{N}_3}{R_f}=\frac{1}{R_f+3+2w_2}.
\end{equation}

%
\section{Conclusions}\label{sec:con}
%

\subsection{Analogies}\label{sec:ana}
It is interesting to compare the results of this work with the results
compiled in \cite{SWM08, BSWM08, BSWM10, JM10,WSJM10, CPNC11} for hard sphere
subject to friction.  There exists a remarkable analogy with similar and
contrasting features between jammed macrostates of disks at criticality along
the dashed arc of our Fig.~\ref{fig:cwlj2-1-f8} and of spheres in the RCP
limit along the vertical borderline of the phase diagram in \cite{SWM08}.  In
both systems the jammed macrostates depend on the compactivity and one
additional control parameter, contributed by gravity in our system and by
friction in the hard-sphere system.

Approaching our critical line via the limit (\ref{eq:18}) with a specific
value for $\Delta$ in (\ref{eq:20}) corresponds to approaching the RCP line
from the left via the limit $X\to0$ for a specific value of $\mu$ (coefficient
of friction) or $Z$ (mechanical coordination number).  The jammed macrostates
to the right of the RCP line are shown in \cite{JM10} to consist of coexisting
domains of RCP disorder and FCC order.  The RCP line can thus be interpreted
as the location of a first-order phase-transition.

In Sec.~\ref{sec:domains} of this work we show indirect evidence that the
macrostates below the (dashed) critical line (see Fig.~\ref{fig:janac1-4-f3})
also consist of coexisting domains of order and disorder. In our case two
distinct types of ordering are explored.  However, we have yet to develop the
mathematical tools for a more direct and quantitative description of the
ordered domains.

There exists a second analogy between the two systems.  It pertains to the
(averaged) mechanical coordination number $Z$.  In the hard-sphere system,
there appears to exist a function $Z(\mu)$, which was shown to vary
monotonically between a low-friction plateau, $Z=6$ and a high-friction
plateau, $Z=4$, for jammed macrostates between the RLP and RCP lines
\cite{WSJM10}.

The confined disks of our system have $Z=3$ with no variation in the absence
of friction.  In Sec.~\ref{sec:fric} we argue that friction stabilizes
additional disk configurations involving disks with only two loaded
contacts. Our argument implies that $Z$ decreases monotonically with
increasing coefficient of friction.  A more quantitative analysis of that
functional relation and its dependence on $\bar{V}$ will have to await an
analysis based on refined mathematical tools.

\subsection{Discussion}\label{sec:dis}
The methodology of this work builds on the foundations of configurational
statistics.  Chief among the assumptions is the existence of jammed
macrostates that are systematically reproducible by some protocol of random
agitations of given intensity and that can be described by few control
variables.

Jammed macrostates of the system under investigation here are parametrized by
two variables reflecting the intensity of the random agitations in two
distinct energy units of relevance.  One unit represents the expansion work of
the piston when it is moved a fraction of one disk diameter.  The other unit
represents the weight of one disk lifted across the channel.

A protocol of random agitations that fits the bill of configurational
statistics for the jammed disks in the narrow channel is not guaranteed to
exist.  If it does exist it may be difficult to implement.  It is no secret
that this is the Achilles' heel of configurational statistics.

The detailed predictions of this work follow rigorously from the assumptions
underlying configurational statistics if friction is absent.  The versatility
of our method of analysis makes it possible to include some dominant effects
of friction in a transparent way.  Our results thus present a situation where
the assumptions of configurational statistics can be put to the test of
simulations and experiments.  The simplicity of the model in conjunction with
the complexity of its behavior is well suited for that purpose.  Any instances
of verification or falsification will shed light on the important question of
how accurately a protocol of random agitations applied to jammed granular
matter can mimic effects of thermal fluctuations.

\subsection{Outlook}\label{sec:out}
Experimental, theoretical, and computational studies of jammed disks are well
established \cite{MSJ+10, LS90, UKW05, ZMSB10} but scenarios with highly
confined geometries have drawn limited attention thus far \cite{BS06, AB09,
  BA11}.  We hope to have shown in this work that such simple systems exhibit
behavior that is worth studying for at least two reasons: there exist
compelling analogies to behavior of more complex systems, exact analytic
results are within reach.

The methodology as presented here is amenable to further development in
several directions: (i) the regime $1+\sqrt{3/4}<H/\sigma<2$ of 32 tiles as
identified in \cite{AB09}, (ii) mixtures of disks with different radii and
masses including rattlers \cite{DW09,GBOS09,ABO+12}, (iii) spatial
correlations of heterogeneities in mass density, and (iv) channels with the
axis oriented vertically.


%
%
\ack
%
Illuminating discussions with S.S. Ashwin are gratefully acknowledged.


\end{document}